\documentclass[twocolumn,english,showpacs,showkeys,preprintnumbers,amsmath,amssymb,prb]{revtex4}
\usepackage[T1]{fontenc}
\usepackage[latin9]{inputenc}
\usepackage{textcomp}
\usepackage{amsmath}
\usepackage{color}
\usepackage{graphicx}
\usepackage{setspace}

\makeatletter

\newcommand{\noun}[1]{\textsc{#1}}
\DeclareRobustCommand{\greektext}{%
 \fontencoding{LGR}\selectfont
 \def\encodingdefault{LGR}}
\DeclareRobustCommand{\textgreek}[1]{\leavevmode{\greektext #1}}
\DeclareFontEncoding{LGR}{}{}

\providecommand{\tabularnewline}{\\}
\makeatother

\usepackage{babel}

\begin{document}

\title{Magnetic Behavior of Single Crystalline Pr$_{5}$Ge$_{3}$\textit{
}and Tb$_{5}$Ge$_{3}$ Compounds}

\author{Devang A. Joshi, A. Thamizhavel and S. K. Dhar }

\affiliation{Department of Condensed Matter Physics and Material Sciences, Tata
Institute of Fundamental Research, Homi Bhaba Road, Colaba, Mumbai
400 005, India.}

\begin{abstract}
\textit{The results of the magnetization studies on Pr$_{5}$Ge$_{3}$and
Tb$_{5}$Ge$_{3}$ single crystals are reported. Single Crystals of
Pr$_{5}$Ge$_{3}$ and Tb$_{5}$Ge$_{3}$ compounds were successfully
grown by Czochralski method. These compounds crystallize in a Mn$_{5}$Si$_{3}$
type hexagonal structure with space group P6$_{3}$/mcm. }\textit{\textcolor{black}{Ferromagnetic
correlations set in at around 36 K in Pr$_{5}$Ge$_{3}$ in the ab
plane followed by an antiferromagnetic transition at 13 K. Along the
c-axis the magnetization shows a ferromagnetic transition around 13
K with an overall ferrimagnetic behavior. }}\textit{At 2K, the magnetic
isotherm of the compound along {[}0001] direction is typical for a
ferromagnet, while a field induced ferromagnetic type response is
observed along the {[}10$\overline{\mathit{1}}$0] direction. Hexagonal
ab plane or {[}10$\overline{\mathit{1}}$0] direction was found to
be the easy axis of magnetization. Tb$_{5}$Ge$_{3}$ orders antiferromagneticaly
at 85 K with the hexagonal ab plane as easy axis of magnetization.
The compound shows a field induced ferromagnetic behavior in its magnetic
isotherm at 2 K.}
\end{abstract}



\maketitle
\begin{singlespace}

\section{\textbf{INTRODUCTION}}
\end{singlespace}

\begin{singlespace}
\textcolor{black}{\noun{R}}\textcolor{black}{$_{5}$}\textcolor{black}{\noun{M}}\textcolor{black}{$_{3}$(R
= rare earths, M = }\textit{\textcolor{black}{p\ }}\textcolor{black}{
block elements) compounds exist for a variety of 'M' elements such
as M = Si, Ge, Ga, Pb and Sn. The crystal structure of these binary
compounds depends upon the atomic sizes of R and M as well as the
type of the }\textit{\textcolor{black}{p}}\textcolor{black}{ \ block
element. Most of }\textcolor{black}{\noun{R}}\textcolor{black}{$_{5}$}\textcolor{black}{\noun{M}}\textcolor{black}{$_{3}$
crystallize in the Mn$_{5}$Si$_{3}$-type hexagonal structure in
which the rare earth atoms occupy two inequivalent crystallographic
4}\textit{\textcolor{black}{d\ }}\textcolor{black}{ and 6}\textit{\textcolor{black}{g}}\textcolor{black}{
\ sites located at (1/3, 2/3, 0) and (x$_{\text{R}}$},0,\textcolor{black}{1/4).
Due to the different near neighbor environment associated with the
two sites, }\textcolor{black}{\noun{R}}\textcolor{black}{$_{5}$}\textcolor{black}{\noun{M}}\textcolor{black}{$_{3}$
compounds typically show complicated magnetic structures despite their
relatively simple formula. A} variety of behaviors such as \textcolor{black}{the}
coexistence of antiferromagnetic and ferromagnetic \textcolor{black}{components,
incommensurate amplitude modulated and conical spin structures and
field induced magnetic configurations are observed. The R$_{5}$Ge$_{3}$
family of compounds was first studied by Buschow and Fast, using polycrystalline
materials\citet{Buschow}}. \textcolor{black}{Bulk magnetization indicates
that}\textcolor{blue}{ }\ Ce$_{5}$Ge$_{3}$ and Nd$_{5}$Ge$_{3}$
\textcolor{black}{are }ferrimagnetic; Pr$_{5}$Ge$_{3}$\textcolor{blue}{,}
Tb$_{5}$Ge$_{3}$, Dy$_{5}$Ge$_{3}$ Ho$_{5}$Ge$_{3}$ and Er$_{5}$Ge$_{3}$
are weak antiferromagnets at low temperature and exhibit a field induced
metamagnetic transition\citet{Buschow}. \textcolor{black}{A neutron
diffraction study on Nd$_{5}$Ge$_{3}$ \citet{Schobinger}was more
revealing and the results could best be explained by assuming a collinear
antiferromagnetic double sheet structure for the Nd atoms at the 6}\textit{\textcolor{black}{g}}\textcolor{black}{
\ site and a canted (or a possible modulated) structure for the 4}\textit{\textcolor{black}{d}}\textcolor{black}{
\ atoms. Below 20 K a fairly strong ferromagnetic component can be
induced by means of an external field in Nd$_{5}$Ge$_{3}$. }
\end{singlespace}

\textcolor{black}{There are very few reports based on single crystals
of R$_{5}$Ge$_{3}$. A single crystal of Ce$_{5}$Ge$_{3}$ was reported
to show dense Kondo behavior \citet{Kurisu}. Tsutaoka et al. have
reported the magnetization and electrical transport properties of
single crystals of Gd$_{5}$Ge$_{3}$ and Tb$_{5}$Ge$_{3}$ \citet{Tsutaoka}.
Gd$_{5}$Ge$_{3}$ undergoes two antiferromagnetic transitions at
76 and 52 K, respectively, while a single antiferromagnetic transition
at 79 K and a large magnetic anisotropy are observed in Tb$_{5}$Ge$_{3}$.
Keeping in mind the complicated magnetic behavior of Nd$_{5}$Ge$_{3}$
in the ordered state, we have been motivated to examine the corresponding
behavior in a single crystal specimen of the neighboring Pr$_{5}$Ge$_{3}$.
The large magnetic anisotropy in Tb$_{5}$Ge$_{3}$ prompted us to
study its magnetization behavior in greater detail than given in \citet{Tsutaoka}.
Accordingly, we have successfully grown single crystals of Pr$_{5}$Ge$_{3}$
and Tb$_{5}$Ge$_{3}$ and the results of our magnetization study
are presented in this report.}

\begin{singlespace}

\section{\textbf{EXPERIMENT}}
\end{singlespace}

\begin{singlespace}
From the Phase diagram of the Pr-Ge and Tb-Ge system, Pr$_{5}$Ge$_{3}$
and Tb$_{5}$Ge$_{3}$ were found to be congruently melting with a
melting point of 1490 $^{\circ}$C and 1900 $^{\circ}$C respectively.
Taking the advantage of this property, we decided to grow both the
single crystals by Czochralski pulling method. Starting materials
were high purity Pr (99.95 \%) and Ge (99.999\%) metals from Leico
industries. Stoichiometric amount of materials were taken to make
a 10 g (polycrystal) melt in a tetra arc furnace. A thin polycrystalline
seed rod of respective compound was immersed into the melt and pulled
at a speed of 10 mm/h in pure and dry argon atmosphere. The grown
single crystal was approximately 2-3 mm in diameter and a photograph
of Pr$_{5}$Ge$_{3}$ pulled single crystal is shown in Fig. 1.
\end{singlespace}

\begin{figure}
\includegraphics[width=0.5\textwidth]{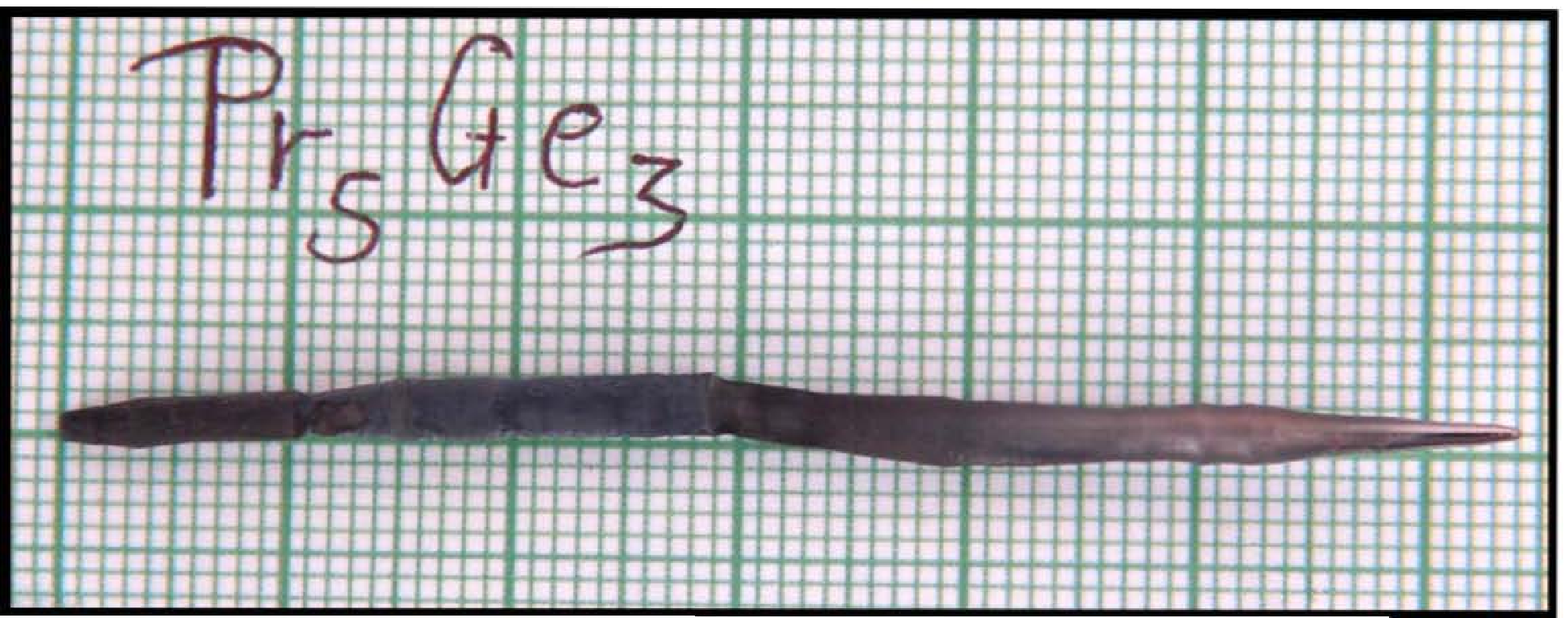}\caption{(Color online) Photograph of pulled Pr$_{5}$Ge$_{3}$ single crystal}

\end{figure}
The phase homogeneity of the crystal was checked using the powder
X-ray diffraction. The single crystals were oriented along the principle
crystallographic directions using back reflection Laue diffraction
method and then cut \textcolor{black}{to required size} for thermal
and magnetic measurements. The magnetic measurements were performed
using superconducting quantum interference device (SQUID - Quantum
Design) and vibrating sample magnetometer (VSM Oxford Instruments),
within a temperature range of 1.8 to 300 K and magnetic fields up
to 120 kOe. The heat capacity measurement was performed using a physical
property measurement system, PPMS (Quantum Design).

\begin{singlespace}
\begin{figure}[h]
\includegraphics[width=0.5\textwidth]{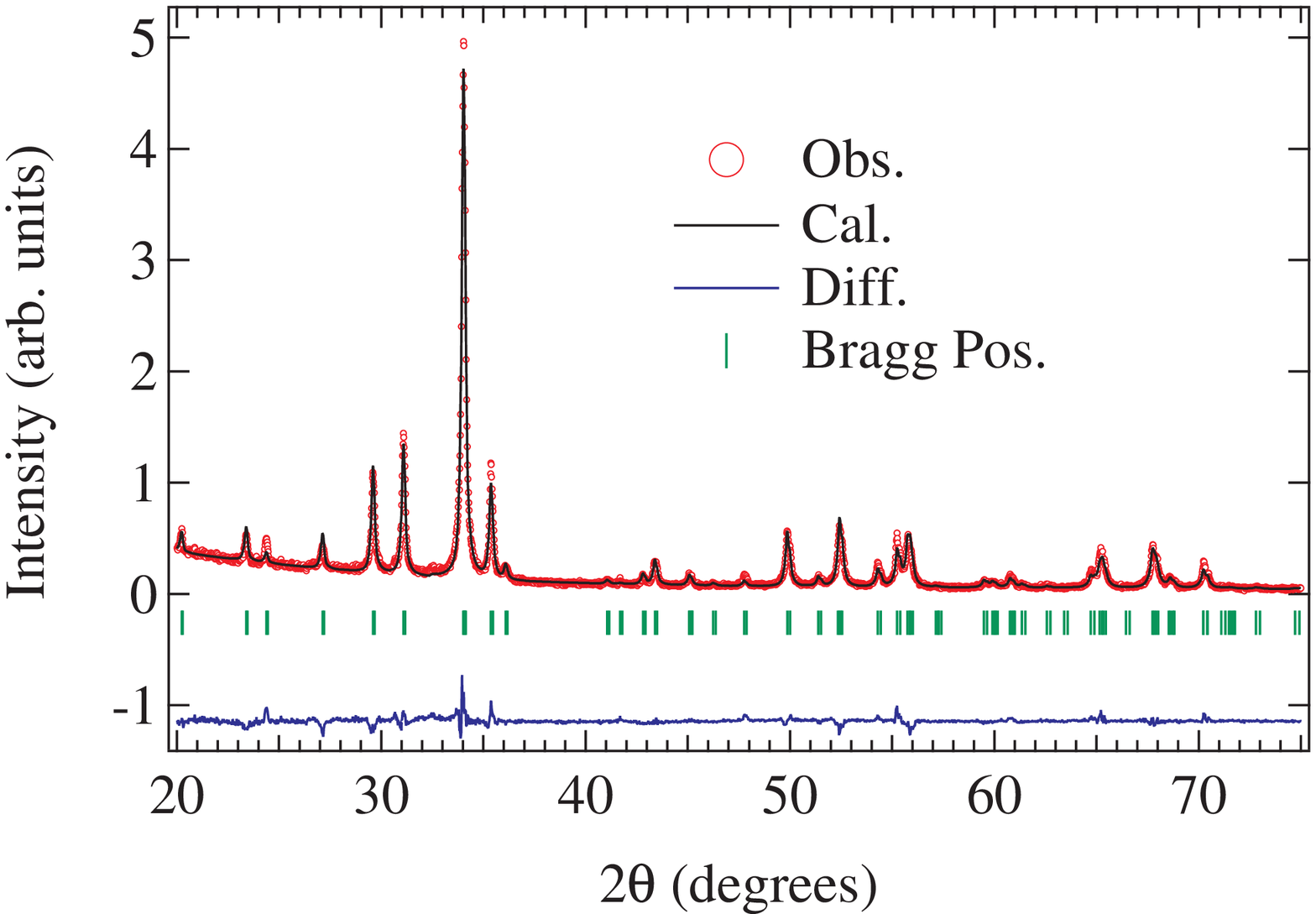}\caption{(Color online) Rietveld analysis of the X-ray powder pattern of Pr$_{5}$Ge$_{3}$
compound. }

\end{figure}

\end{singlespace}

\begin{singlespace}

\section{\textbf{RESULT}}
\end{singlespace}

\begin{singlespace}
\textcolor{black}{As mentioned in introduction, both Pr$_{5}$G}e$_{3}$
and Tb$_{5}$Ge$_{3}$ form in Mn$_{5}$Si$_{3}$ type hexagonal structure
with space group P6$_{3}$/mcm (No. 193). In order to confirm the
phase homogeneity of the compounds with proper \textcolor{black}{crystallographic
and latti}ce parameters, a Rietveld analysis of the observed X-ray
pattern was done using FULLPROF program as shown in Fig. 2. \textcolor{black}{The
single phase nature of the samples was also confirmed by scanning
electron microscopy (SEM).}\textcolor{magenta}{ }The lattice parameters
obtained from the Rietveld analysis for Pr$_{5}$Ge$_{3}$ and Tb$_{5}$Ge$_{3}$
are \textit{a} = 8.804 $\textrm{\AA}$ ; \textit{c} = 6.588 $\textrm{\AA}$
and \textit{a} = 8.474 $\textrm{\AA}$ ; \textit{c} = 6.305 $\textrm{\AA}$
respectively. The lattice parameters are in close agreement with the
reported ones \citet{Buschow}. The refined crystallographic parameters
for Pr$_{5}$Ge$_{3}$ are presented in Table 1 and the crystal structure
is shown in Fig. 3. The black line edges represent a unit cell consisting
of two formula units of Pr$_{5}$Ge$_{3}$. The \textit{xy} planes
on top, bottom and \textcolor{black}{the}\textcolor{blue}{ }\ middle
which consist of only Pr atoms labeled as Pr1 represent the crystallographic
4\textit{d} planes. The remaining two planes which consist of both
Pr (labeled as Pr2) and Ge atoms represent the 6\textit{g} plane.
The crystal structure can be viewed as staking of two different \textit{xy}
planes (planes containing the 4\textit{d} and 6\textit{g} sites) alternately
staked along the '\textit{c}' axis. The rare earth atoms have different
occupancies at 4\textit{d} and 6\textit{g} crystallographic sites.
\textcolor{black}{The nearest neighbor 4}\textit{\textcolor{black}{d-}}\textcolor{black}{4}\textit{\textcolor{black}{d}}\textcolor{black}{
\ interatomic distance is 3.343 $\textrm{\AA}$ which is appreciably
shorter than that corresponding}\textit{\textcolor{black}{ }}\textcolor{black}{\ 6}\textit{\textcolor{black}{g-}}\textcolor{black}{6}\textit{\textcolor{black}{g}}\textcolor{black}{
\ atomic distance of 4.005 $\textrm{\AA}$\citet{Buschow}. The interlayer
(4}\textit{\textcolor{black}{d-}}\textcolor{black}{6}\textit{\textcolor{black}{g}}\textcolor{black}{)
atomic distance is at an intermediate value of 3.758$\textrm{\AA}$}\textit{\textcolor{black}{\citet{Buschow}.}}
\end{singlespace}

\begin{figure}
\includegraphics[width=0.5\textwidth]{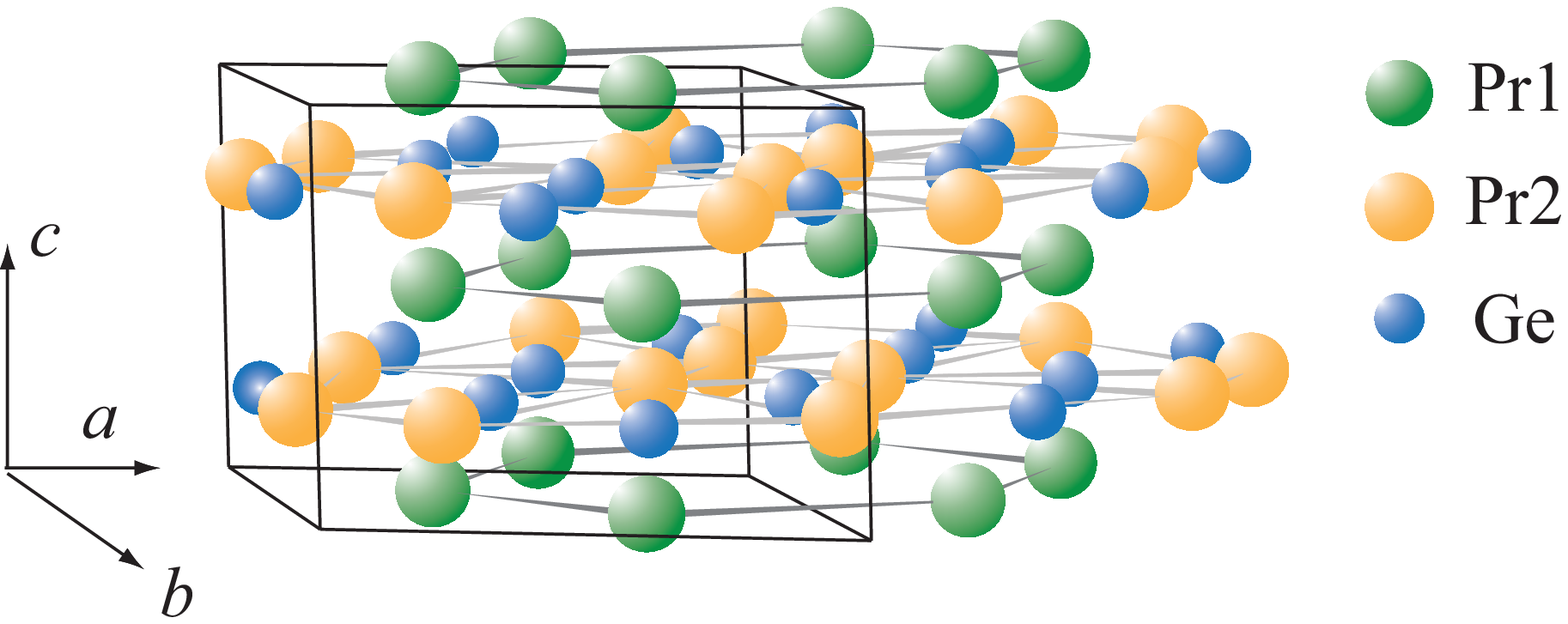}\caption{(Color online) Crystal structure of Pr$_{5}$Ge$_{3}$ compound, the
black lined edges represent the unit cell. The top, bottom and the
central planes containing only Pr1 along the \textit{c} axis represent
the 4\textit{d} planes and the remaining planes containing Pr2 and
Ge represent the \textcolor{black}{6}\textit{g} planes. }

\end{figure}
\begin{table}[h]
\begin{tabular}{ccccccc}
 &  &  &  &  &  & \tabularnewline
\hline
\hline 
Atom & Site  & x & y & z & U$_{{\rm eq}}$($\textrm{\AA}^{2}$) & Occ.\tabularnewline
 & Symmetry &  &  &  &  & \tabularnewline
\hline
Pr1 & 4d & 0.333 & 0.666 & 0.000 & 0.269(2) & 1\tabularnewline
Pr2 & 6g & 0.230(1) & 0.000 & 0.250 & 2.218(4) & 1.5\tabularnewline
Ge & 6g & 0.604(6) & 0.000 & 0.250 & 0.051(1) & 1.5\tabularnewline
\hline
\hline 
 &  &  &  &  &  & \tabularnewline
\end{tabular}\caption{Refined crystallographic parameters for Pr$_{5}$Ge$_{3}$}

\end{table}
\begin{figure}
\includegraphics[width=0.5\textwidth]{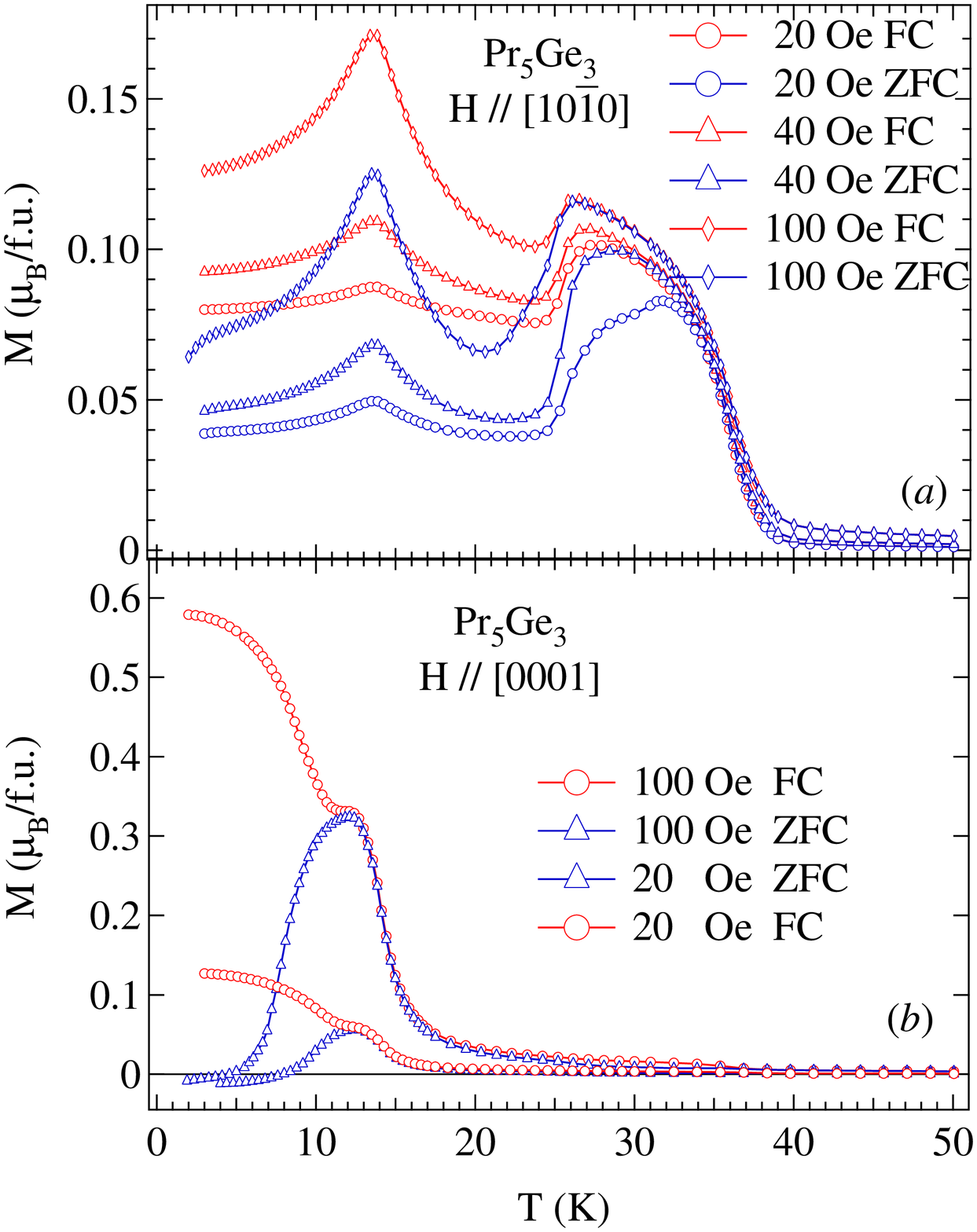}\caption{(Color online)\textbf{ a:} Magnetization \textit{\textcolor{black}{vs}}.
temperature curves for Pr$_{5}$Ge$_{3}$ with field parallel to {[}10$\overline{1}$0].
\textbf{b:} \textcolor{black}{Similar} curves with field parallel
to {[}0001]. }

\end{figure}

\begin{singlespace}

\subsection{Pr$_{5}$Ge$_{3}$}
\end{singlespace}

\begin{singlespace}
\textcolor{black}{Fig. 4a shows the magnetization }\textit{\textcolor{black}{vs}}\textcolor{black}{
\ temperature curve for Pr$_{5}$Ge$_{3}$ under zero field cooled
(ZFC) and field cooled (FC) conditions with applied low fields of
20 , 40 and 100 Oe along {[}10$\overline{1}$0] direction. Signatures
of two magnetic transitions at T$_{1}$ $\approx$ 13 K and T$_{2}$}
\textcolor{black}{$\approx$ 36 K are seen in all the three fields.
The upturn in the magnetization near T$_{2}$ suggests the onset of
ferromagnetic correlations in the }\textit{\textcolor{black}{ab}}\textcolor{black}{
\ plane}\textcolor{blue}{, }\textcolor{black}{while the transition
at 13 K appears to have an antiferromagnetic character. Its position
is independent of the applied field while the peak at higher temperature
(corresponding to T2 transition) broadens and shifts considerably
to lower temperatures as the field is increased from 20 to 100 Oe.
Thermomagnetic irreversibility under ZFC and FC conditions is observed
for the transition at 36 K, which typically occurs in ferromagnets
with large magnetocrystalline anisotropy. The latter also explains
the peak shift to lower temperatures with increasing field as arising
from an interplay of the applied and coercive field. Between T$_{1}$
and T$_{2}$ the FC magnetization decreases in a limited range as
the temperature is decreased at all fields, which is different from
the typical saturation behavior of magnetization observed in ferromagnet
in the FC mode. }
\end{singlespace}

\textcolor{black}{The magnetization with field parallel to {[}0001]
axis is shown in Fig. 4b. There is only one magnetic transition at
$\approx$ 13 K in contrast to that with field parallel to {[}10$\overline{1}$0].
The nature of the transition seems to be ferromagnetic but the negative
magnetization in the ZFC condition indicates an overall ferrimagnetic
type of behavior. In the above measurements we have ensured that the
notional zero fields in which the sample was cooled is not negative
such that the negative ZFC magnetization is not attributed to a large
coercive field. This transition seems to be the analog of that appearing
along {[}10$\overline{1}$0] direction at $\approx$13 K but the behavior
is totally different. The high field (3 kOe) susceptibility curve
is shown in Fig. 5a. Along {[}10$\overline{1}$0] direction the T$_{2}$
peak has broadened appreciably to the point of becoming almost imperceptible;
however, the low temperature peak T$_{1}$ is clearly seen. On the
other hand the magnetization along {[}0001] direction shows a ferromagnetic
ordering.} %
\begin{figure}
\includegraphics[width=0.5\textwidth]{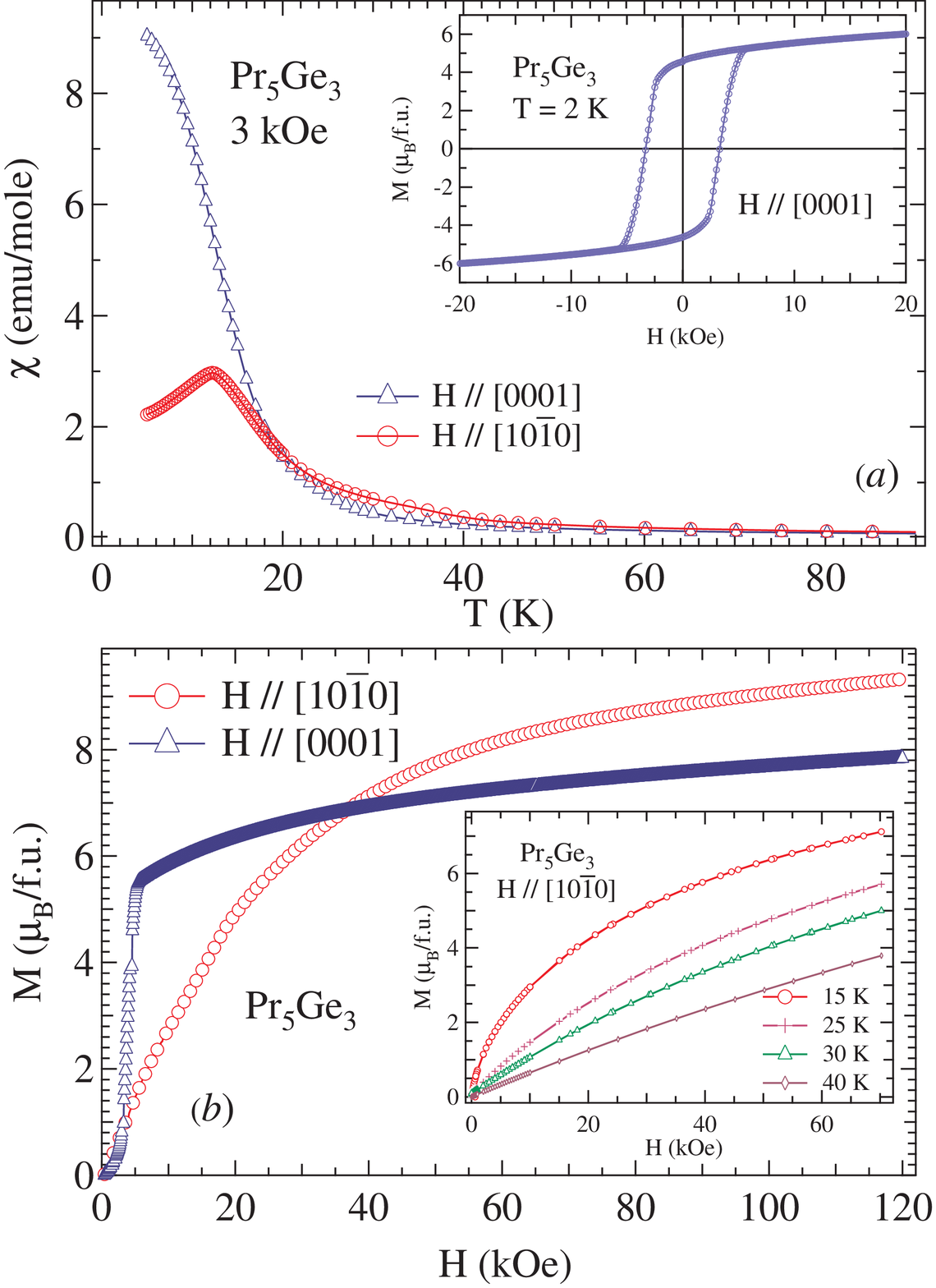}\caption{(Color online)\textbf{\textcolor{black}{ a:}}\textcolor{black}{ \ High
field susceptibility of Pr$_{5}$Ge$_{3}$ along {[}10$\overline{1}$0]
and {[}0001] directions with inset showing the magnetic isotherm at
2 K with field along {[}0001] direction. }\textbf{\textcolor{black}{b:}}\textcolor{black}{
\ Magnetic isotherms for Pr$_{5}$Ge$_{3}$ at 2 K along both the
two directions; the inset shows the temperature variation of magnetization
along {[}10$\overline{1}$0] direction.}}

\end{figure}
\textcolor{black}{The magnetic isotherms at 2 K for Pr$_{5}$Ge$_{3}$
with field along {[}0001] and {[}10$\overline{1}$0] are shown in
Fig.5b with the inset showing the magnetic isotherms at 40 K, 30 K,
25 K and 15 K along the {[}10$\overline{1}$0] direction. The plots
with the field along {[}10$\overline{1}$0] show that overall the
magnetization increases as the temperature is decreased attaining
substantial values at high fields. Overall the behavior for T $\leq$
30 K appears to be a superposition of both ferromagnetic and antiferromagnetic
components. In particular focussing our attention on the 2 K plot
the magnetization initially increases linearly with field and then
moves towards saturation at high fields. The magnetization at low
fields is less then that along {[}0001] direction, but it overtakes
the latter at \ensuremath{\approx} 37 kOe and remains higher up to
the highest applied field of 120 kOe. The magnetization at 120 kOe
is \ensuremath{\approx} 9.3 $\mu$$_{B}$/ f.u. In the reverse direction,
the magnetization exhibits a hysteresis with a coercive field of \ensuremath{\approx}
1 kOe (not shown). Along {[}0001] direction the magnetization increases
sharply with field as expected for a ferromagnetic compound and exhibits
a hysteresis with a coercive field of 3.3 kOe as shown in the inset
of Fig 5a. The magnetization at 120 kOe is \ensuremath{\approx} 7.}8
\textcolor{black}{$\mu$$_{B}$/} f.u. %
\begin{figure}
\includegraphics[width=0.5\textwidth]{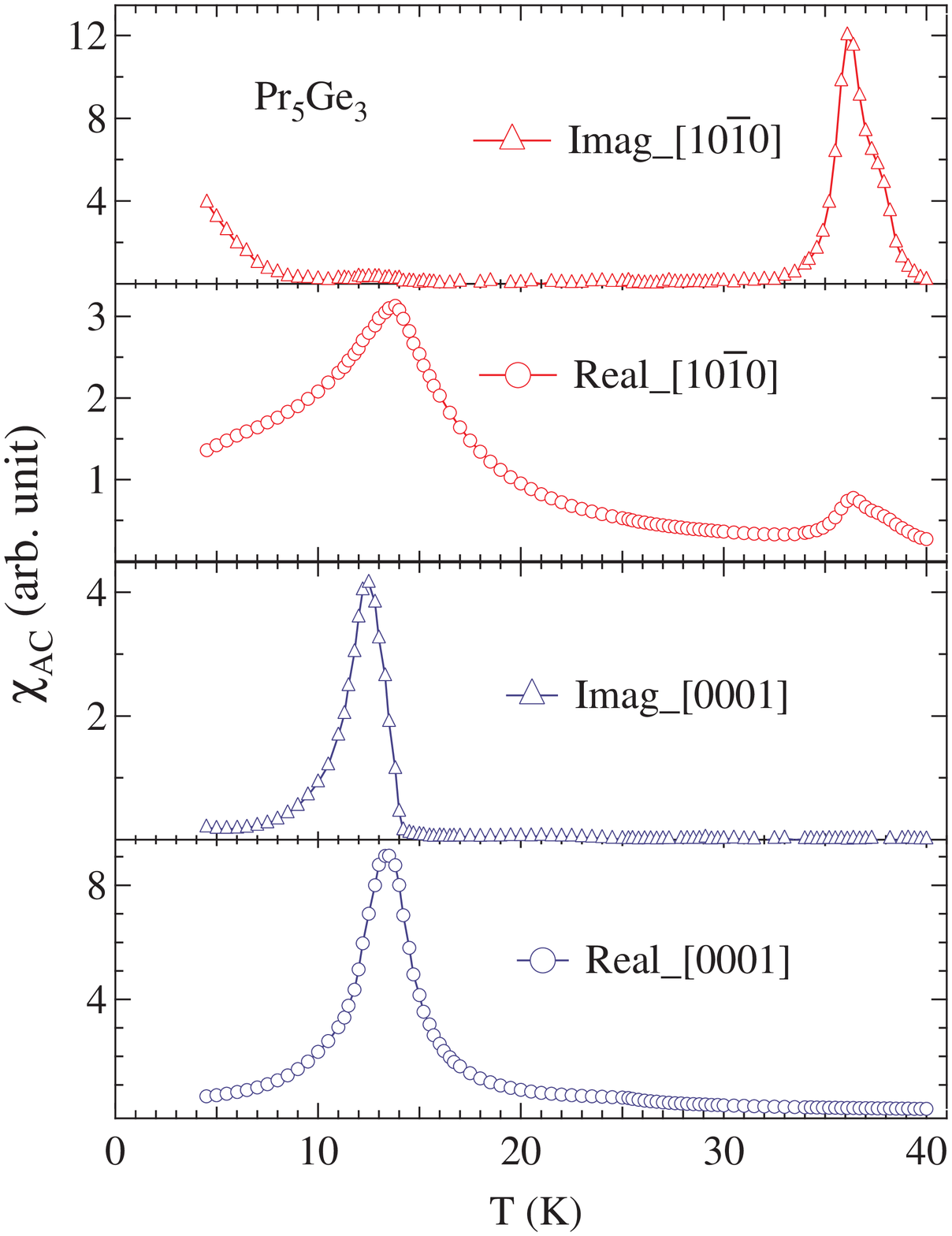}\caption{(Color online) AC susceptibility of Pr$_{5}$Ge$_{3}$ with AC field
along {[}10$\overline{1}$0] and {[}0001] directions.}

\end{figure}

\begin{singlespace}
\textcolor{black}{In order to get more information on the complex
magnetic phenomenon revealed by DC magnetization data as presented
above, we measured the AC susceptibility of the compound with AC field
applied along the two crystallographic directions {[}10$\overline{1}$0]
and {[}0001], respectively as shown in Fig. 6. When the AC field is
parallel to {[}10$\overline{1}$0] direction, the real part of AC
susceptibility (\textgreek{q}\textasciiacute{}) shows peaks at approximately
36 and 13 K, whereas the imaginary part of AC susceptibility (\textgreek{q}\H{ })
shows a peak only at 36 K. A peak in \textgreek{q}\textasciiacute{},
reflects any type of magnetic ordering, whereas a peak in \textgreek{q}\H{ }
appears only if a ferromagnetic component is present. Hence a collinear
antiferromagnetic ordering will not reflect in the imaginary part
of AC susceptibility}. The presence of peak in both \textgreek{q}\textasciiacute{}
and \textgreek{q}\H{ } at \textcolor{black}{36 K indicates a magnetic
ordering with net ferromagnetic component, whereas the absence of
peak in \textgreek{q}\H{ } at 13 K indicates a collinear type of antiferromagnetic
ordering. This also supports our DC magnetization results where magnetization
under FC and ZFC condition bifurcates only at the T$_{2}$ tr}ansition.
\textcolor{black}{It may be noted that the peak at 13 K in \textgreek{q}\textasciiacute{}
is stronger than that at the higher temperature.}\textcolor{blue}{
}\ The increase in \textgreek{q}\H{ } at low temperatures \textcolor{black}{below
approximately 8 K is presently not understood. When the AC field is
applied along {[}0001] direction, both the real and imaginary parts
show peak at approximately 13 K, consistent with the DC magnetization
results, which show the dominant ferromagnetic behavior of the compound
along this direction.} %
\begin{figure}
\includegraphics[width=0.5\textwidth]{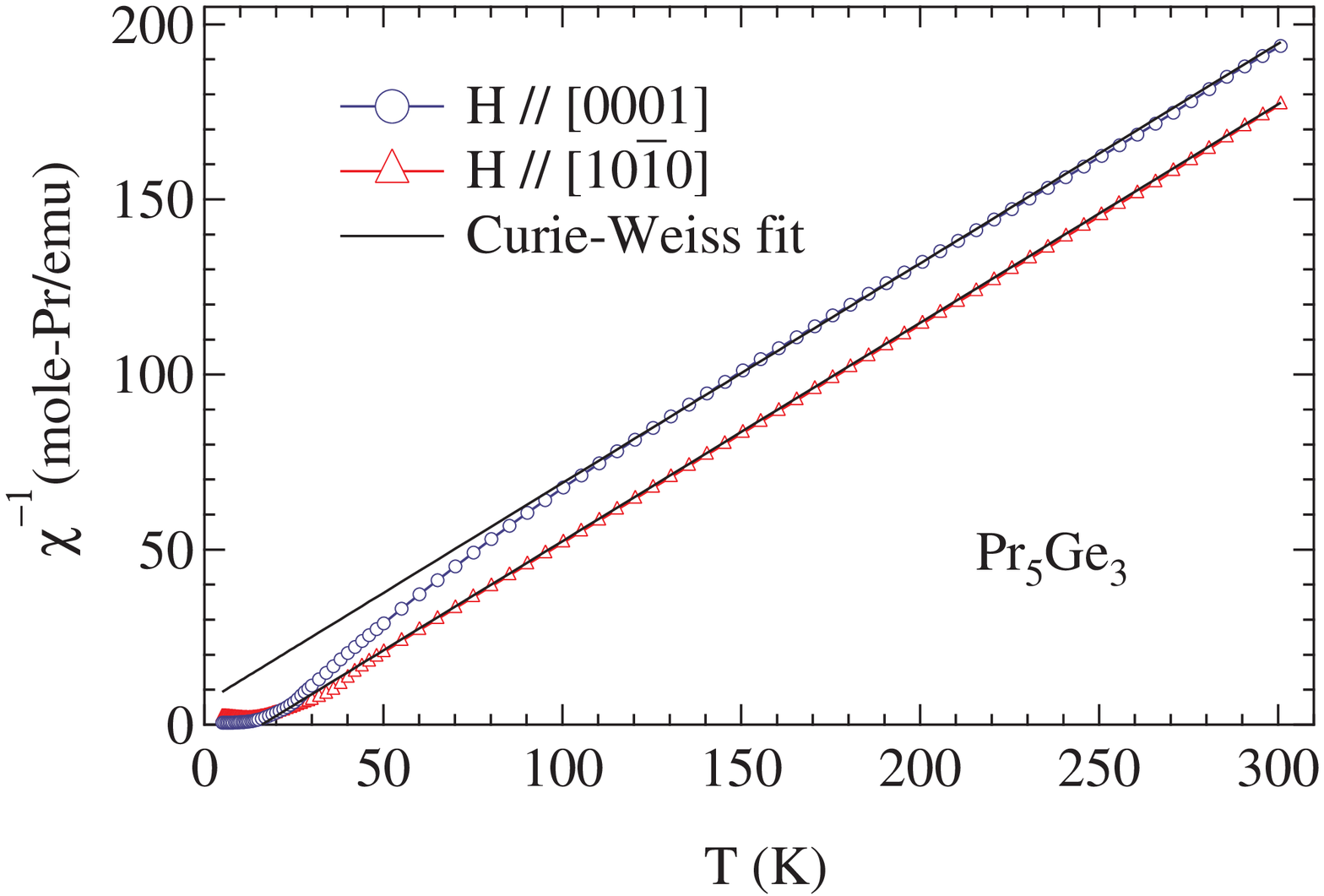}

\caption{(Color online) Inverse susceptibility of Pr$_{5}$Ge$_{3}$ with a
Curie-Weiss fit.}

\end{figure}

\textcolor{black}{The inverse susceptibility of the compound along
the two directions is shown in Fig. 7. The solid lines are fits of
the Curie-Weiss law to the data and furnish $\theta$$_{P}$ = 16
K, $\mu$$_{eff}$ = 3.58 $\mu$$_{B}$ and $\theta$$_{P}$ = -10
K, $\mu$$_{eff}$ = 3.58 $\mu$$_{B}$ along {[}10$\overline{1}$0]
and {[}0001] directions respectively.}\textcolor{magenta}{ }\textcolor{black}{\ In
the paramagnetic state the susceptibility along {[}10$\overline{1}$0]
direction is higher than that along {[}0001] at high temperatures
but at low temperature the susceptibility along {[}0001] is higher
because of the ferromagnetic type of ordering (Fig. 5a). It must be
noted that the inverse susceptibility along {[}0001] direction deviates
from the Curie-Weiss fit at much higher temperature (100 K) than along
{[}10$\overline{1}$0] direction indicating a dominant crystal field
effect along the '}\textit{\textcolor{black}{c}}\textcolor{black}{'
axis of magnetization. The high temperature susceptibility and the
magnetic isotherms at 2 K show that the easy axis of magnetization
in Pr$_{5}$Ge$_{3}$ is along the }\textit{\textcolor{black}{ab}}\textcolor{black}{
\ plane.}
\end{singlespace}

\begin{figure}
\includegraphics[width=0.5\textwidth]{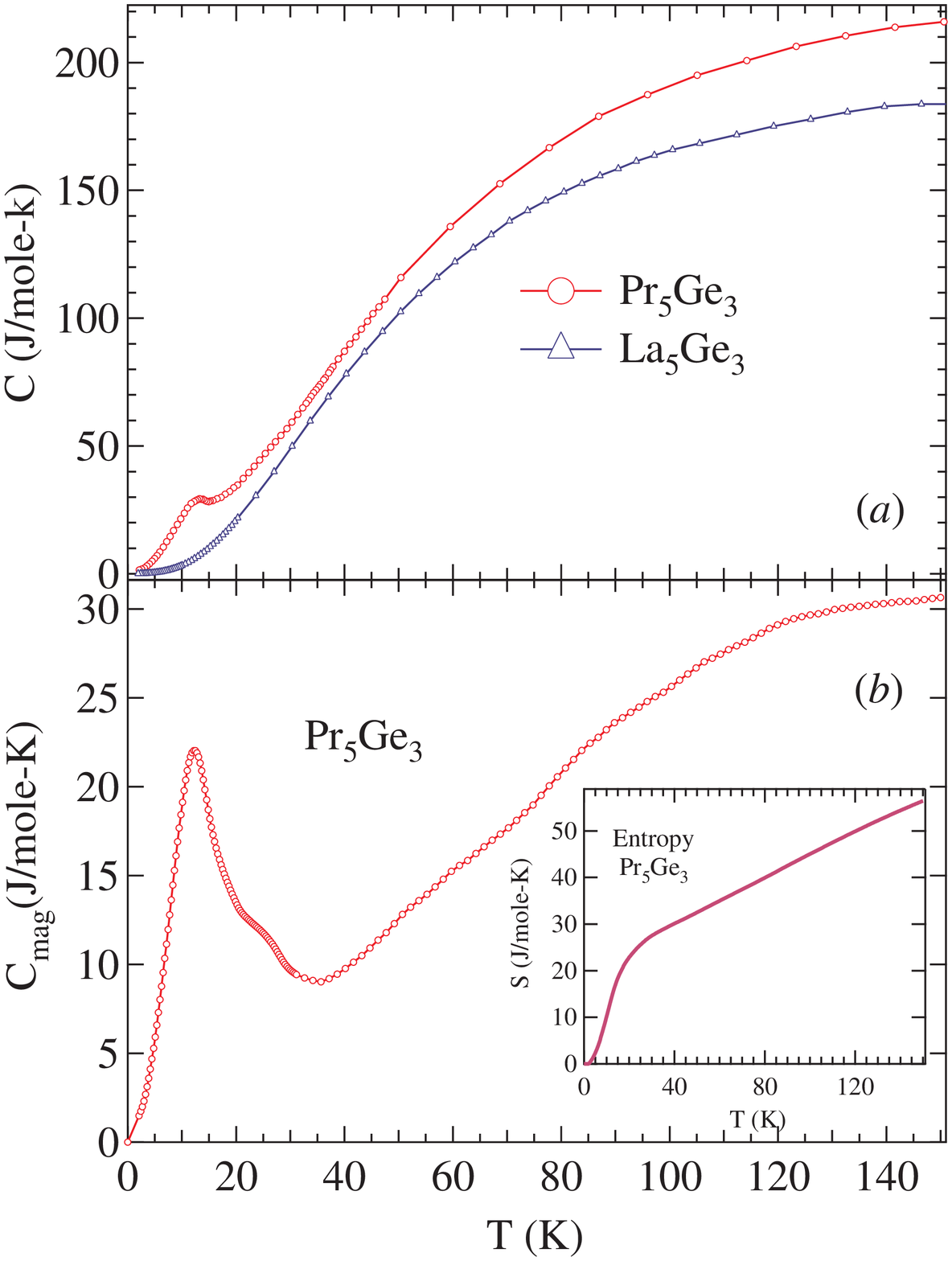}\caption{(Color online)\textbf{\textcolor{black}{ a:}} \textcolor{black}{Comparison
of the heat capacity of Pr$_{5}$Ge$_{3}$ and La$_{5}$Ge$_{3}$.
}\textbf{\textcolor{black}{b:}}\textcolor{black}{ the 4}\textit{\textcolor{black}{f}}\textcolor{black}{
\ contribution to the heat capacity of Pr$_{5}$Ge$_{3}$ with the
inset showing the calculated entropy.}}

\end{figure}
\textcolor{black}{The heat capacity of Pr$_{5}$Ge$_{3}$ and the
non-magnetic polycrystalline reference compound La$_{5}$Ge$_{3}$
is depicted in Fig.8a. The heat capacity of La$_{5}$Ge$_{3}$ increases
monotonically with temperature as expected for a nonmagnetic compound.
The heat capacity of Pr$_{5}$Ge$_{3}$ shows a minor peak at $\approx$
13 K and then increases with increase in temperature. Above 50 K the
difference between the heat capacity of Pr$_{5}$Ge$_{3}$ and La$_{5}$Ge$_{3}$
increases for which a likely reason could be the presence of Schottky
contribution arising from the Boltzmann fractional occupation of the
thermally excited crystal electric field split levels in Pr$_{5}$Ge$_{3}$
or La$_{5}$Ge$_{3}$ may cease to be a good reference for the lattice
heat capacity at high temperatures. Some evidence for the former comes
from the fact that the high temperature part (above 100 K) of the
heat capacity of Pr$_{5}$Ge$_{3}$ could not be fitted to the sum
of electronic ($\gamma T$) and lattice contributions (Debye Integral)
alone. The 4}\textit{\textcolor{black}{f}}\textcolor{black}{ \ contribution
to the heat capacity (C$_{4f}$) of Pr$_{5}$Ge$_{3}$ (Fig.8b) was
isolated by subtracting the heat capacity of La$_{5}$Ge$_{3}$ taking
into account the slightly differing atomic masses of La and Pr. The
peak at 13 K is now sharper in agreement with the magnetic ordering
along both the crystallographic axes as deduced from the magnetization
data above. There is no apparent anomaly at 36 K but a broad hump
centered at $\approx$ 25 K is seen. We believe that the magnetic
contribution to the heat capacity around 36 K may get submerged under
the over-riding Schottky anomaly which appears to be present as inferred
from the upturn in the 4}\textit{\textcolor{black}{f}}\textcolor{black}{
\ heat capacity beginning at T $\approx$ 35 K. The entropy calculated
from C$_{4f}$ is 3.2 and 6.3 J/mol-Pr K at 13 and 40 K, respectively,
compared to the value 5.76 J/mol-Pr K for a doublet ground state.
This shows that a substantial short range order exists above 13 K.
The T$_{2}$ peak in the magnetization along {[}10$\overline{1}$0]
and the hump in the heat capacity may be a signature of the short
range order. }

\begin{figure}[h]
\includegraphics[width=0.5\textwidth]{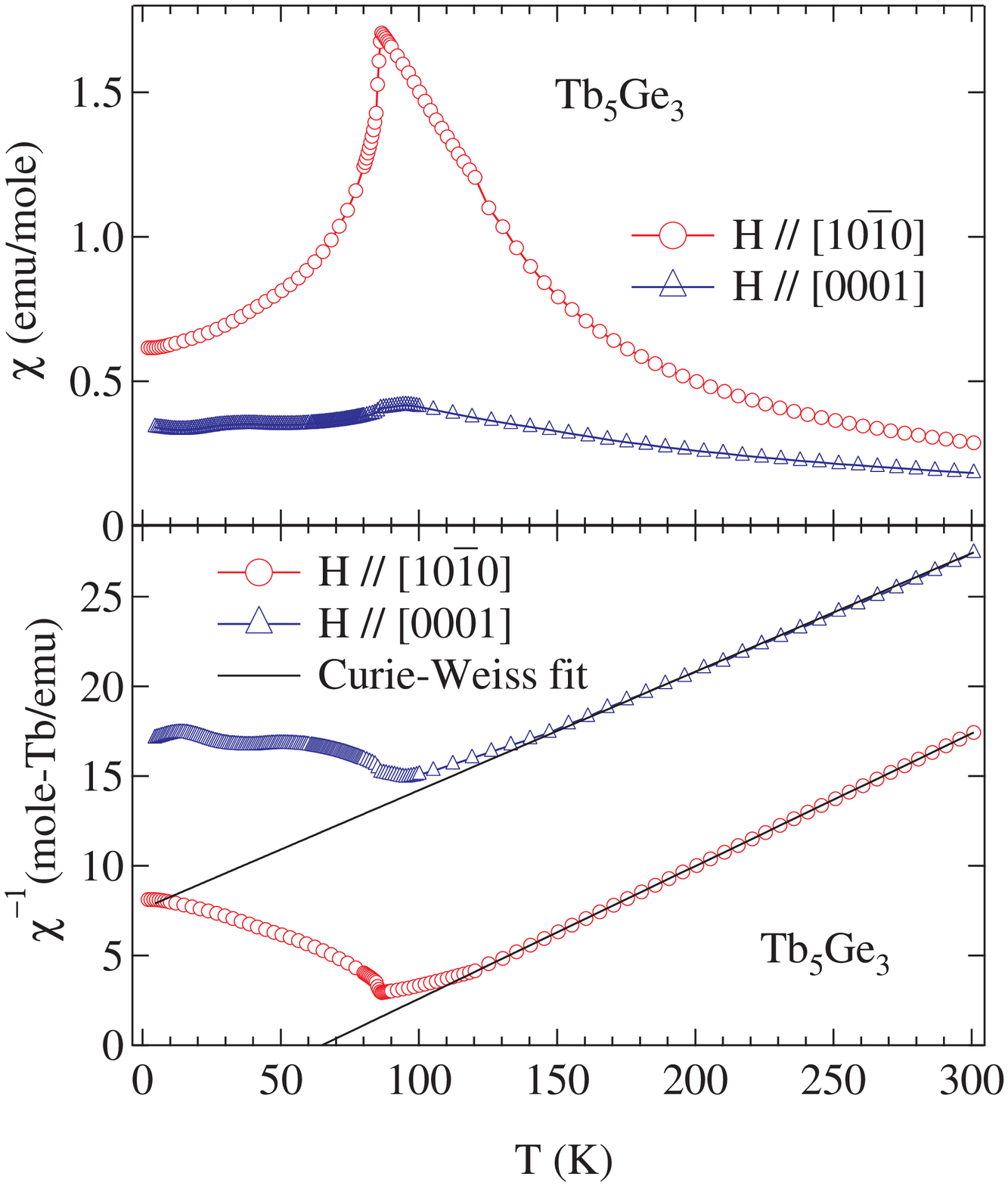}\caption{(Color online)\textbf{ a:} Magnetic susceptibility of Tb$_{5}$Ge$_{3}$
along both the crystallographic direction. \textbf{b:} Inverse magnetic
susceptibility; the solid line through the data point indicate the
Curie-Weiss fit. }

\end{figure}

\subsection{Tb$_{5}$Ge$_{3}$}

\begin{singlespace}
\textcolor{black}{The susceptibility of Tb$_{5}$Ge$_{3}$ with field
applied along the crystallographic directions is shown in Fig. 9a.
The susceptibility with field parallel to {[}10$\overline{1}$0] direction
exhibits a sharp peak at T$_{N}$ = 85 K characteristic of an antiferromagnetic
transition. The susceptibility with field parallel to {[}0001] shows
a kink at \ensuremath{\approx} 85 K followed by a broad hump at low
temperatures (between 20 to 50 K), indicating a relatively complex
behavior. The inverse susceptibility is plotted in Fig. 9b. In the
paramagnetic region the fit of the Curie-Weiss law to the data furnishes
$\theta$$_{P}$ = 62 K, $\mu$$_{eff}$ = 9.75 $\mu$$_{B}$ and
$\theta$$_{P}$ = -140 K, $\mu$$_{eff}$ = 9.6 $\mu$$_{B}$ along
{[}10$\overline{1}$0] and {[}0001] directions respectively. The magnitude
of the susceptibility along the two axes reflects the highly anisotropic
magnetic response and also shows that {[}10$\overline{1}$0] is the
easy direction of magnetization.}
\end{singlespace}

\begin{figure}
\includegraphics[width=0.5\textwidth]{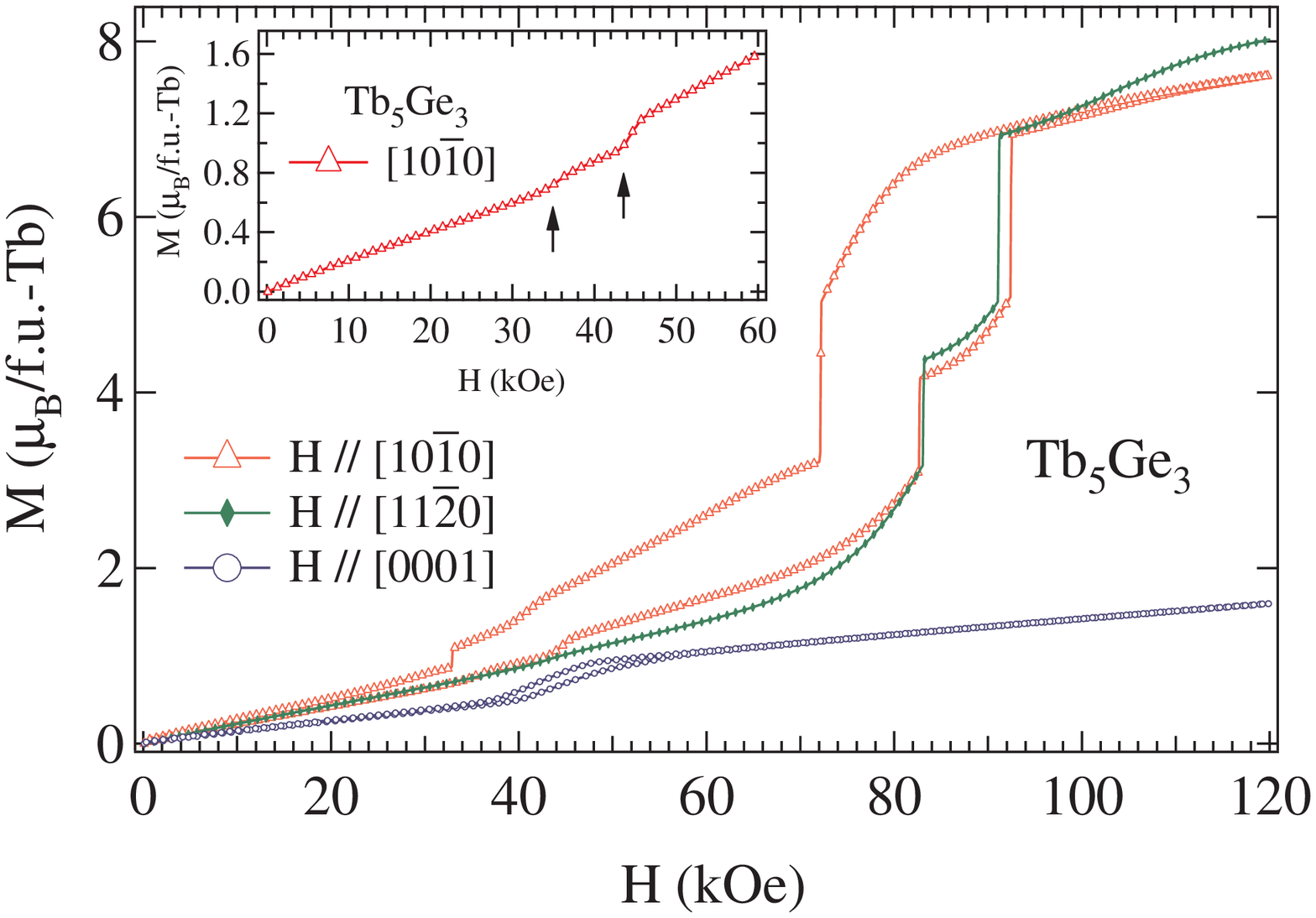}

\caption{(Color online)\textcolor{black}{ Magnetic isotherm of Tb$_{5}$Ge$_{3}$
with field along {[}10$\overline{1}$0], {[}0001] and {[}11$\overline{2}$0]
directions. The inset shows the expanded magnetic isotherm at low
field with arrows pointing the metamagnetic transitions.}}

\end{figure}

\begin{singlespace}
\textcolor{black}{Fig. 10 shows the magnetic isotherm of Tb$_{5}$Ge$_{3}$
with field along {[}10$\overline{1}$0], {[}0001] and {[}11$\overline{2}$0]
(lies within the }\textit{\textcolor{black}{ab}}\textcolor{black}{
\ plane) directions at 2 K. We have applied fields up to 120 kOe
which exceed significantly the maximum applied field of 50 kOe used
by Tsuoka et al. As a result we see extra features in the magnetization
at high fields. The magnetization along {[}10$\overline{1}$0] direction
undergoes multiple metamagnetic transitions at \ensuremath{\approx}
34, 42, 82 and 92 kOe, respectively. The former two magnetic transitions
are spin-flip type where as the latter two are spin-flop type. The
curved nature of the magnetic isotherm between approximately 60 and
80 kOe in the increasing direction of the field suggests a canted
configuration of the antiferromagnetic state (AF II, Fig. 11). The
maximum magnetization at 120 kOe is \ensuremath{\approx} 7.5 $\mu$$_{B}$/Tb,
which is little less than the saturation moment of Tb$^{3+}$ion.
Hence the latter two metamagnetic transitions drive the compound to
the field induced ferromagnetic state. A significant amount of hysteresis
is observed during the demagnetization of the sample, due to the pinning
of the domain walls in an anisotropic ferromagnetic material. The
field induced ferromagnetic state of the compound indicates the weakly
coupled antiferromagnetic nature of the compound. The magnetization
along the hard axis undergoes a minor metamagnetic transition \ensuremath{\approx}
40 kOe. This result is in contrast with that reported previously \citet{Tsutaoka},
where no metamagnetic transition is encountered up to a field of 50
kOe. The magnetization at 120 kOe and 2 K is \ensuremath{\approx}
1.8 $\mu$$_{B}$/Tb, which is far less compared to that obtained
with field along {[}10$\overline{1}$0] direction, as expected for
a hard axis of magnetization. When the field is applied along the
{[}11$\overline{2}$0] direction the magnetization undergoes two spin-flop
type metamagnetic transitions at approximately the same values as
that with field along {[}10$\overline{1}$0] direction. The magnetization
at 120 kOe and 2 K is \ensuremath{\approx} 8 $\mu$$_{B}$/Tb, which
is \ensuremath{\approx} 0.5 $\mu$$_{B}$/Tb higher then that along
{[}10$\overline{1}$0] direction and close to the saturation moment
of the Tb$^{3+}$ion. The magnetic phase diagram of the compound constructed
from the temperature variation of the magnetic isotherms (not shown)
with field applied along the easy axis of magnetization {[}10$\overline{1}$0]
is shown in Fig. 11. The symbols AF, AF I, AF II and AF III represents
the three different antiferromagnetic states (including the canted
antiferromagnetic state) of the compound.}
\end{singlespace}

\begin{figure}
\includegraphics[width=0.5\textwidth]{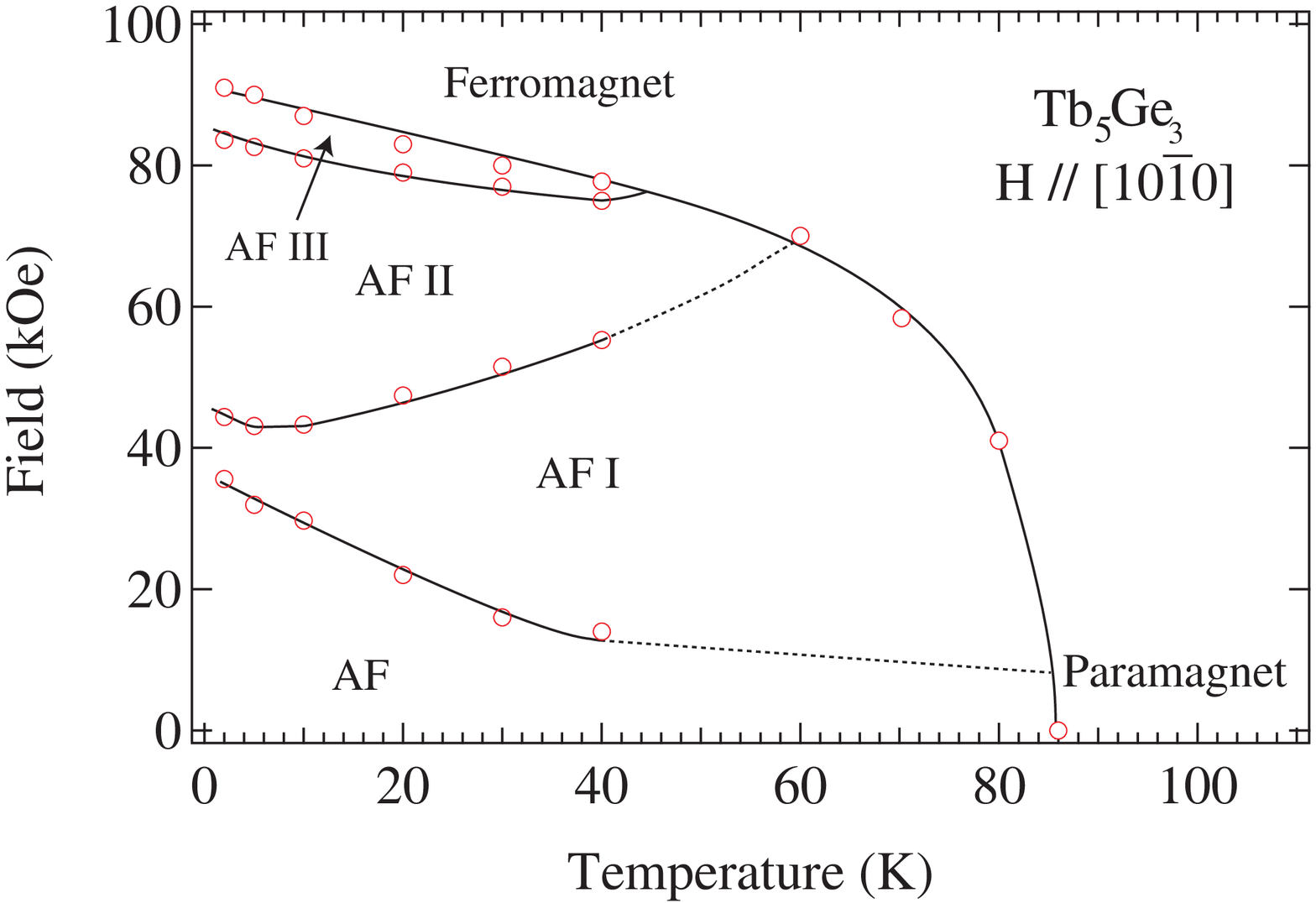}\caption{(Color online) Magnetic phase diagram of Tb$_{5}$Ge$_{3}$ constructed
using the temperature variation of magnetic isotherms with field applied
along {[}10$\overline{1}$0] direction. \textcolor{black}{The dotted
lines represents the expected imaginary path of the curve. }}

\end{figure}
\medskip{}

\section{DISCUSSION}

\textcolor{black}{In the preceding sections we have presented the
interesting magnetic behavior of Pr$_{5}$Ge$_{3}$ and Tb$_{5}$Ge$_{3}$.
Pr$_{5}$Ge$_{3}$ orders ferrimagneticaly near 13 K with a dominant
ferromagnetic component when the field is applied along {[}0001] direction.
Along the {[}10$\overline{1}$0] direction there are two transitions
at T$_{1}$ and T$_{2}$; the transition at T$_{1}$ appears to be
antiferromagnetic, while the upturn occurring at the higher temperature
T$_{2}$ suggests the presence of ferromagnetic type correlations.
At high fields the behavior of 2 K isotherms along both the crystallographic
axes is dominantly ferromagnetic type. Since there are two symmetry
inequivalent crystallographic sites, 4}\textit{\textcolor{black}{d}}\textcolor{black}{
\ and 6}\textit{\textcolor{black}{g}}\textcolor{black}{, for the
rare earth ion and the nearest neighbor 4}\textit{\textcolor{black}{d}}\textcolor{black}{-4}\textit{\textcolor{black}{d}}\textcolor{black}{
\ distance is significantly lesser than the corresponding 6}\textit{\textcolor{black}{g}}\textcolor{black}{-6}\textit{\textcolor{black}{g
}}\textcolor{black}{distance, site dependent magnetic response is
in principle possible. The latter is actually seen in the neighboring
(similar lattice parameters) isostructural compound Nd$_{5}$Ge$_{3}$
as mentioned in the Introduction. In Pr$_{5}$Ge$_{3}$ the behavior
of the magnetization around \ensuremath{\approx} 36 K with field along
{[}10$\overline{1}$0] direction, which bears the signature of the
onset of the ferromagnetic type correlations can plausibly be attributed
to the ions present at the }\textit{\textcolor{black}{4d}}\textcolor{black}{
\ site. The 4}\textit{\textcolor{black}{d}}\textcolor{black}{ \ site
moments being relatively closer to each other compared to the other
ions are}\textcolor{magenta}{ }\textcolor{black}{coupled more strongly
by the indirect RKKY exchange interaction. With further decrease in
the temperature the second magnetic transition at \ensuremath{\approx}
13 K may be due to the }\textit{\textcolor{black}{ab}}\textcolor{black}{
\ plane-projected collinear antiferromagnetic type ordering of the
Pr$^{3+}$ moments at the }\textit{\textcolor{black}{6g}}\textcolor{black}{
\ site. As already mentioned above, the latter is strongly suggested
by AC susceptibility. The \textgreek{q}\textasciiacute{} (real part)
of the AC susceptibility includes contributions from both magnetic
rotation and domain wall movement, whereas \textgreek{q}\H{ } (imaginary
part) reflects the energy loss due to the movement of domain walls.
If the antiferromagnetic ordering is collinear then the resulting
moment is zero and hence no movement of domain walls is involved,
resulting in the absence of peak in \textgreek{q}\H{ }. We further
speculate that between T$_{1}$ and T$_{2}$, the evolution of 4}\textit{\textcolor{black}{d}}\textcolor{black}{
-4}\textit{\textcolor{black}{d}}\textcolor{black}{, 4}\textit{\textcolor{black}{d}}\textcolor{black}{
-6}\textit{\textcolor{black}{g}}\textcolor{black}{ \ and 6}\textit{\textcolor{black}{g}}\textcolor{black}{
-6}\textit{\textcolor{black}{g}}\textcolor{black}{ \ interactions
with temperature is such that it overall gives rise to a decrease
in the FC magnetization in a limited range of temperature, as already
mentioned above. }

\textcolor{black}{Along {[}0001] direction, the peak seen in both
the real and the imaginary part of the AC susceptibility supports
the ferromagnetic nature of the transition at 13 K, in tune with the
DC magnetization results presented above. The relatively sharp nature
of the peak is typically seen in highly anisotropic magnetic compounds,
corroborated by the high coercivity (\ensuremath{\approx} 3 kOe at
2 K) seen in the magnetic isotherm. }

\textcolor{black}{The magnetic isotherms along {[}10$\overline{1}$0]
and {[}0001] directions at 2 K show that at high fields Pr$_{5}$Ge$_{3}$
is like a ferromagnet. At 2 K the moment at 120 kOe is \ensuremath{\approx}
9.3 $\mu$$_{B}$/ f.u. and 7.8 $\mu$$_{B}$/ f.u. along these two
directions, repectively. Keeping in mind that the saturation moment
of free Pr$^{3+}$ ion is 3.2 $\mu$$_{B}$ and the likely reduction
in the moment due to the crystal electric fields, the magnetization
response involves the polarization of Pr ions of both sublattices.
Of course, the magnetization along {[}10$\overline{1}$0] is larger
because it is the easy axis of magnetization. }

\textcolor{black}{The absence of any anomaly at 36 K when the field
is applied along the {[}0001] axis may be rationalized by assuming
that the easy axis of magnetization for the Pr-4}\textit{\textcolor{black}{d}}\textcolor{black}{
\ moments is in the }\textit{\textcolor{black}{ab}}\textcolor{black}{-plane.
As the temperature is lowered a dominant ferromagnetic component is
observed at $\approx$ 13 K with an overall ferrimagnetic behavior.
Since the ordering temperature ($\approx$ 13 K) is similar to that
occurring also along {[}10$\overline{1}$0] direction, it is possible
that the orientation of the 6}\textit{\textcolor{black}{g}}\textcolor{black}{
\ moments is such that it can be resolved into a collinear antiferromagnetic
configuration in the }\textit{\textcolor{black}{ab}}\textcolor{black}{
\ plane and a ferromagnetic component along the }\textit{\textcolor{black}{c}}\textcolor{black}{-axis.}\textcolor{magenta}{
}\textcolor{black}{\ In order to explain the ferrimagnetic response
at low temperatures, we postulate that i) the anisotropy of the 6}\textit{\textcolor{black}{g}}\textcolor{black}{
\ sublattice is stronger than that of 4}\textit{\textcolor{black}{d}}\textcolor{black}{-sublattice
and ii) the exchange interaction between the 6}\textit{\textcolor{black}{g}}\textcolor{black}{
\ and the 4}\textit{\textcolor{black}{d}}\textcolor{black}{ \ moments
forces the latter to reorient from their easy axis. }\textcolor{magenta}{ }\textcolor{black}{In
order to get the full details of the magnetic configuration, neutron
diffraction on a single crystal is required. The arrangement of the
magnetic moments in these compounds can be very complex. For example.
a recent neutron diffraction experiment on Ho$_{5}$Ge$_{3}$\citet{Morozkin}
finds between T$\mathrm{_{N1}}$ = 27 K and T$\mathrm{_{N2}}$ = 18
K, a sine-modulated ordering with two propagation vectors K$_{1}$
= {[}0,0,$\pm$3/10] and K$_{2}$ = {[}0. 1/2,0]. The magnetic configuration
changes below T$\mathrm{_{N2}}$ and is described by four propagation
vectors K$_{1}$ = {[}0,0, $\pm$ 3/10], K$_{2}$ = {[}0,1/2,0], K$_{3}$
= {[}0,0, $\pm$ 2/5] and K$_{4}$ = {[}$\pm$1/5, $\pm$1/5, 0].}\textcolor{magenta}{ }

\begin{singlespace}
\textcolor{black}{Tb$_{5}$Ge$_{3}$, by contrast, shows a relatively
simpler process of magnetic ordering. A peak at T$_{N}$= 85 K is
seen in both directions, though it is far more prominent along {[}10$\overline{1}$0]
(ab plane), the easy axis of magnetization. The leveling of the susceptibility
at low temperatures along {[}10$\overline{1}$0] direction may be
due to the incommensurate magnetic transition reported by \citet{Schobinger1}to
occur between 75 to 50 K, which transforms into a spiral type antiferromagnetic
structure at low temperatures. The compound is a weakly coupled antiferromagnet
in the sense that polycrystalline average of the paramagnetic Curie
temperature is low (-16 K) and also the compound undergoes a field
induced ferromagnetic transition at 2 K. The saturation moment obtained
at 120 kOe and 2 K is close to that reported from the neutron diffraction
\citet{Schobinger1}for the Tb$^{3+}$ ion in Tb$_{5}$Ge$_{3}$.
Along the hard axis of magnetization there is an evidence of a metamagnetic
transition at \ensuremath{\approx} 40 kOe. Such transitions along
the hard axis of magnetization are generally attributed to the crossing
over of the crystal field split energy levels as shown in case of
NdRhIn$_{5}$ \citet{Hieu}. The reason for hysteresis appearing in
the magnetic isotherm is not known.}
\end{singlespace}

\section{CONCLUSIONS}

\noindent In conclusion, we have studied the magnetic properties of
single crystalline Pr$_{5}$Ge$_{3}$ and Tb$_{5}$Ge$_{3}$. Hexagonal
\textit{\textcolor{black}{ab}} plane or {[}10$\overline{1}$0] direction
was found to be the easy axis of magnetization for Pr$_{5}$Ge$_{3}$.
In the \textit{\textcolor{black}{ab}} plane, the magnetization shows
a ferromagnetic type upturn at \ensuremath{\approx} 36 K followed
by a collinear antiferromagnetic ordering of the moments at \ensuremath{\approx}
13 K. Along {[}0001] direction the compound shows a dominant ferromagnetic
transition at \ensuremath{\approx} 13 K with an overall ferrimagnetic
type behavior. At 2 K, the magnetic isotherm of the compound along
{[}0001] direction is typical for a ferromagnet, while a field induced
ferromagnetic type response is observed along the {[}10$\overline{1}$0]
direction. Tb$_{5}$Ge$_{3}$ was found to order antiferromagnetically
at 85 K with a hexagonal \textit{ab} \ plane as a easy axis of magnetization.
The compound undergo a field induced ferromagnetic state at low temperatures
along the easy axis of magnetization.

\end{document}